\documentstyle[aps,manuscript,epsf]{revtex} 

\begin{document}
\title{\bf 
%
                    Calculation of the transition matrix and 
                      of the occupation probabilities 
                               for the states
                         of the Oslo sandpile model
}
\author{
\'Alvaro Corral%
\cite{email} 
}
\address{
Departament de F\'\i sica, 
Universitat Aut\`onoma de Barcelona,
Edifici Cc, E-08193 Bellaterra, Barcelona, Spain 
}
%
\date{\today}

\maketitle 
%
%
\begin{abstract}
The Oslo sandpile model, or if one wants to be precise, ricepile model,
is a cellular automaton designed to model experiments on granular piles
displaying self-organized criticality.
We present an analytic treatment that allows the calculation of the transition
probabilities between the different configurations 
of the system;
from here, using the theory of Markov chains,
we can obtain the stationary occupation distribution,
which tell us that the phase space is occupied
with probabilities that vary in many orders of magnitude
from one state to another.
Our results show how the complexity of this simple model 
is built as the number of elements increases, and allows,
for a given system size, the exact calculation
of the avalanche size distribution and other properties
related to the profile of the pile.
%
%
\end{abstract}
%

\pacs{
PACS numbers:  
89.75.-k  
89.75.Da  
05.65.+b,    
45.70.Ht,   
}
%

\narrowtext
\setcounter{page}{1} 
%

\section{ Introduction and definition of the Oslo model }

Self-organized criticality (SOC),
born from the deep insights of Bak {\it et al.},
deals with the emergence of scale invariance
in slowly-driven nonequilibrium 
systems \cite{Bak,Jensen}.
The phenomenon is illustrated with the archetypical example of a pile of sand, 
and realized in computer simulations of diverse sandpile models,
which are mainly based on the original Bak-Tang-Wiesenfeld (BTW) model \cite{BTW}.
However, the relevance of SOC for real granular matter
was unclear until Frette {\it et al.} \cite{Frette} 
performed experiments on a $1+1-$dimensional pile of rice;
these experiments and some others \cite{Christensen,Malthe} were modeled
with a cellular automaton introduced in Ref. \cite{Christensen}, 
later called the Oslo model.

The Oslo model has the interest of being (as far as we know) the first
SOC sandpile model or, more appropriately, ricepile model, able to reproduce
experimental results.
For the avalanche properties \cite{Frette}, the concordance with experiments is only
qualitative, whereas for the transport of individual grains \cite{Christensen}, 
and for the surface roughness \cite{Malthe}, 
the agreement is also quantitative.
Moreover, the Oslo model is remarkable as a simple model of SOC, 
because it displays this nontrivial behavior in one dimension.

The model is designed to mimic the experimental situation in Refs. 
\cite{Frette,Christensen,Malthe}: 
grains are slowly added at a fixed position 
on a quasi one-dimensional substrate 
which is in between two parallel vertical plates;
just at the (let us say) left of the position of addition
a wall prevents the falling of the grains;
on the other side, the right boundary is open. 
The model assumes a discrete space, $x=1,2\dots L$, from left to right,
as well as discrete time and field (the height of the pile, or number of grains). 
The grains pile up
in columns until the local slope somewhere is too large, 
then the upper grain becomes unstable
and is transferred to the next column to the right (from $x$ to $x+1$). 
This transference can induce further instabilities and therefore a chain reaction
or avalanche. 
A slow-driving condition imposes that during the avalanches the external
addition of grains is interrupted;
this implies that the evolution of the model (and of the experiments)  
takes place in two separated time scales:
a slow time scale for the grain addition
and a fast time scale for the evolution of the avalanches.

In terms of the height of the $x$ column, $h(x)$, and the local slope,
defined as $z(x) \equiv h(x)-h(x+1)$, taking $h(L+1)\equiv 0$,
these prescriptions are expressed in the following rules \cite{Christensen}: 
\begin{equation}
\begin{array}{lllllll}
      \mbox{if } z(x) \le z_{th}(x) \ \forall x  \Rightarrow 
       h(1) \rightarrow h(1)+1, \\
      \mbox{if }  z(x) > z_{th}(x) 
      \Rightarrow  
      \left\{ \begin{array}{llll}
         h(x) \rightarrow h(x)-1,\\
         h(x+1) \rightarrow h(x+1)+1,\\
         z_{th}(x) \rightarrow rand \\ 
      \end{array} \right. \\
\end{array}
\label{rules_rice_pile0}
\end{equation}
(where the update is supposed to take place in parallel).
Here $z_{th}(x)$ refers to a local threshold, 
which rather than being constant changes
with every toppling at $x$ to a random value $rand$, chosen as
\begin{equation}
rand = \left\{
\begin{array}{l}
1 \mbox{ with probability } p, \\
2 \mbox{ with probability } q=1-p. \\
\end{array}
\right.
\end{equation}

These simple rules, Eqs. (1) and (2), 
together with the boundary condition, $h(L+1)\equiv 0$,
completely define the Oslo model.
Nevertheless, it is convenient to express rule (1)
in terms only of the slope, turning out to be,
\begin{equation}
\begin{array}{lllllll}
      \mbox{if } z(x) \le z_{th}(x) \ \forall x  \Rightarrow 
       z(1) \rightarrow z(1)+1, \\
      \mbox{if }  z(x) > z_{th}(x) 
      \Rightarrow  \left\{ \begin{array}{llll}
      z(x) \rightarrow z(x)-2,\\
      z(x-1) \rightarrow z(x-1)+1,\\
      z(x+1) \rightarrow z(x+1)+1,\\
      z_{th}(x) \rightarrow rand, \\ 
\end{array} \right. \\
\mbox{for } x=1,2 \dots L-1 \mbox{ and}
      \\
      \mbox{if }  z(L) > z_{th}(L)
      \Rightarrow  \left\{ \begin{array}{lll}
      z(L) \rightarrow z(L)-1,\\
      z(L-1) \rightarrow z(L-1)+1,\\
      z_{th}(L) \rightarrow rand \\
\end{array} \right.
\end{array}
\label{rules_rice_pile}
\end{equation}
(taking into account that at $x=0$ the $z$ variable is not defined).
Notice how the dynamics of the grains has allowed to define a different
dynamics for the units of slope, which can be considered as some kind of
virtual particles. 
Both dynamics are conservative, except at the boundaries,
which curiously are reversed: 
the open end for the grains at $x=L$ is a closed boundary
for the slopes and vice versa at $x=1$. 
It is important to have this in mind to avoid confusions.

The Oslo model is essentially the one-dimensional BTW model, 
but with fixed addition at $x=1$ and an open boundary condition
for the grains at $x=L$.
The key different ingredient, which makes the model critical in one dimension,
is the selection of dynamically changing thresholds, 
to account for the heterogeneities of a real system.
In this way, whereas the randomness in the BTW sandpile is external,
in the Oslo model it is internal, as in real ricepile experiments.
This spirit is original from the philosophy of Ref. \cite{Frette93},
although the model there is much more complicated.

Even with the simplicity of its definition, Eqs. (1) and (2),
the Oslo model gives rise to an astonishing complex behavior.
As it is usual in SOC systems, it shows a power-law distribution
of avalanche sizes \cite{Christensen,Amaral} and avalanche
durations \cite{Amaral}, signaling the existence of no
characteristic scales for the avalanche process.
But also it has been shown that the transport of the grains
through the pile is anomalous, in the sense that there is no
normal diffusion but a power-law distributed transit time \cite{Christensen},
spanning many orders of magnitude
(just as it happens in the experiments, as we have mentioned).
This has been explained by the fact that the time that a grain
is trapped at a given position is also broadly distributed \cite{Boguna},
which has in its turn being related to a kind of skewed fractional
Brownian motion for the variations of the height,
in the antipersistent case \cite{Hopcraft2003}.
Moreover, the distances traveled by the grains during an avalanche
turn out to be L\'evy flights \cite{Boguna}, i.e., again scale free,
despite the nearest-neighbor rules.
Additionally, the time fluctuations of the profile scale 
with a roughness exponent that is in good concordance
with the experiments \cite{Christensen,Malthe}.
In fact, the exponents of all previous magnitudes can be related
by several scaling laws.

Further, the time sequence of the transit times shows a clear multifractal
spectrum \cite{Pastor}, whereas the time sequence of the mean slope
displays $1/f$ noise \cite{Zhang}, in contrast to the BTW model \cite{Jensen}.
The model also allows one to study the transition from intermittent
behavior to continuous flow, just increasing the driving rate 
and breaking the time-scale separation \cite{Corral99}.
Very recently it has been shown that a small damage performed in the system
does not spread, and therefore the sensitivity of the model to the initial
conditions is quite different to that of chaotic systems and to
what was expected from systems ``at the edge of chaos''
\cite{Stapleton}.
On the other hand, there exists an exact mapping between this model
an a model of interface depinning, which establishes the existence of 
a wide universality class for these non-equilibrium systems
related to the quenched Edwards-Wilkinson equation 
\cite{Paczuski,Nakanishi,Vieira,Pruessner}.
In a previous paper, 
we have also signaled the similarities between the Oslo model and
the recurrence of real earthquakes \cite{Corral2003}.
For all these reasons we can consider the Oslo model 
as the analog of the Ising model for slowly driven complex systems.

In spite of these fascinating properties, our understanding of the Oslo 
model comes mainly from computer simulations and some scaling arguments;
no analytical solution exists nor seems possible in the near future
(as it is the general case for nonequilibrium systems).
Hence, the exact enumeration of the number
of states in the attractor for this model, performed by Chua and Christensen,
is very remarkable \cite{Chua}. 
They found that this number increases exponentially with system size as
\begin{equation}
{N}_A = \frac{a G^L+ a^{-1} G^{-L}}{\sqrt{5}},
\label{Na}
\end{equation}
with $G$ the golden mean, 
$G=(3+\sqrt{5})/2 \simeq 2.6$, and $a=(1+\sqrt{5})/2 \simeq 1.6$.

In this article, we are going to derive exact expressions for the transition
probabilities between states in the Oslo model; 
using the results for a system of size $L-1$ we will get these probabilities
in a system of size $L$. 
The increase in just one unit of the size of the system
leads to an enormous increase in the complexity of the resulting equations;
we have a kind of machine for building complexity.
With the transition probabilities it is possible to obtain the stationary
occupation distribution, 
which is the probability with which a state 
is visited in the asymptotic regime.
We will get that, in contrast to other SOC models,
the states in the attractor are not equally likely;
rather, the range of occupation probabilities varies
dramatically in many orders of magnitude.
Once the stationary occupation distribution is known, several other quantities, 
directly related with the profile of the pile,
as the mean-slope distribution and the avalanche-size distribution
can be obtained.
(We have very recently become aware that D. Dhar has undertaken an
analysis of precisely the same problem \cite{Dhar2003};
nevertheless, his approach is entirely different to ours
and both works can be considered as complementary of each other.)

%
\section{Some properties of the model}
%

Two types of states, or configurations, are possible in the system:
unstable states, with at least one local slope value above 
its local threshold, $z(x)>z_{th}(x)$,
and metastable states, where all the slopes are below threshold,
$z(x) \le z_{th}(x) \ \forall x$.
Unstable states evolve by means of avalanches towards 
metastable states,
but the addition of a new grain can make metastable configurations
become unstable again, and so on.

We are only interested on metastable states.
If we assume that the initial slopes $z(x)$ are not negative
\cite{negative},
the possible stable values for this variable are 0,1, and 2
(values $\ge 3$ are always unstable);
therefore, in a system of size $L$ a metastable configuration will be fully
specified by an $L$-vector whose components are the slope values
$0,1$, or $2$.
For instance, we can consider a state $s$ to be $s=\{210\dots 1\}$
which means $z(1)=2,z(2)=1,z(3)=0$, and so on up to $z(L)=1$
(alternatively, the metastable state $s$ can be viewed as an integer
with $L$ digits expressed in base 3).
Notice that we do not need to keep track of the thresholds $z_{th}(x)$;
this is so because the dynamics can be described in an alternative
but equivalent way, using the following rule:
\begin{equation}
\mbox{if } z(x) \mbox{ has just changed its value to: } \\
\left\{
\begin{array}{l}
1, \mbox{ or less, no toppling, }\\
2, \mbox{ toppling with probability } p, \\
3, \mbox{ or greater, toppling.}
\end{array}
\right.
\label{alternative}
\end{equation}
We stress that this rule has to be applied only when 
the site at $x$ has received one (or more) units of slope
or has toppled in the previous avalanche (fast) time step.
(This works because we have only defined two possible values
for the threshold, with three values the situation would be
more delicate.)

A very useful property in order to study the evolution of the system
will be the Abelian symmetry, first considered in sandpile models 
by Dhar \cite{Dhar}.
It states that the order in which 
units of slope are added and sites over threshold topple
does not matter for the final configuration of the pile;
therefore, we will be allowed to topple the sites in the most convenient
sequence to keep the process manageable in the calculations.
The demonstration of this property in our case is similar 
to that in Refs. \cite{Dhar,Dhar99,Dhar99b}
but taking into account that we have evolving thresholds.
If we consider two sites $x$ and $y$ that are unstable it is easy to see
that we get the same state no matter which one topples first,
since after the toppling of $x$, site $y$ will still remain unstable,
and the quantity any toppling site is reset
and the quantity transferred to the neighbors will be the same,
independently of the order.
The same reasoning can be extended to more than two over-threshold sites.
But this is so only if
the random thresholds for sites $x$ and $y$
are equally chosen in each possible sequence of topplings,
that is, we need a predefined sequence of thresholds at each site,
or, from a computational point of view, 
instead of having a single random number generator,
a different generator must be used for each site.
From a similar reasoning as before, the addition of grains (or slope) at $x=1$
commutes with the toppling of any unstable site.

On the other hand,
the evolution of the pile can be described by means of a finite Markov chain;
indeed, the probability that a given state transforms into another state
depends only on the two states, and not on the previous history,
thanks to rule (4).
In particular, 
the probability that a metastable state $i$ evolves to a
new metastable state $j$ after the addition of one grain 
and at the end of the corresponding avalanche
is independent on the previous states of the pile,
and can be obtained by means of the unstable states
that separate the states $i$ and $j$.
These probabilities constitute a matrix 
that will be referred to as $\cal W$, with elements $W_{i j}$.
Probability theory imposes that $W_{ij} \ge 0$ 
and that the files of the matrix are normalized to one
(in the probabilistic sense), i.e., $\sum_j W_{ij}=1$.
A matrix with this properties constitutes a 
stochastic matrix.
(Further, as $\cal W$ is constant through the time evolution,
we are dealing with a homogeneous Markov chain.)

\section{Characterization of the attractor and 
existence of a unique stationary distribution}

Some results from the theory of Markov chains can be applied at this point.
To get the stationary properties of the pile it will be crucial to
have a well defined stationary occupation distribution;
this quantity gives the probability with which every state is visited 
in the asymptotic regime, that is, in the attractor,
and it is represented by a vector where each component
corresponds to a configuration of the system.
The stationary occupation distribution is simply referred to 
as the stationary distribution in Markov-chain theory,
but here we are interested in many other probability distributions
in the stationary case, as for instance that for the avalanche sizes.
(Another common name is ergodic distribution.)

For completeness, let us explain that the attractor can be defined
as the set of recurrent, or persistent states, being these 
the states for which the return probability is exactly one.
More precisely, if a state is visited at some time, 
there is a probability one that it will be visited again in the future.
In contrast, for transient states this probability is smaller
than one, or even zero.

At this point it is convenient to use graph theory to represent a Markov chain:
the transition probability matrix $\cal W$ 
defines a graph where nodes $i$ and $j$ are directly connected
if $W_{i j} \ne 0$, otherwise, there is no direct link between $i$ and $j$;
that is, we have the graph of the possible transitions in one (slow) time step.

The existence of a unique stationary occupation distribution,
independent on the initial conditions,
is guaranteed if the graph associated to the matrix $\cal W$ 
has only one non-periodic final class \cite{Gordon},
(this is also a necessary condition).
A final class is a strongly connected component whose elements
have no transitions to elements outside the class.
A strongly connected component is a part of the graph in which
any pair of nodes, or states, can be connected in both directions;
in other words, $i$ and $f$ stay in a circuit and 
it is possible to reach state $f$ starting from $i$
and vice versa.
A final class represents then nothing else than an attractor 
in which the system can settle after a transient period. 
The periodicity of a strongly connected component is the greatest common divisor
of the length of their circuits;
if this number is one the graph is non-periodic.
This is for instance the case in the presence of loops 
(circuits with just one element, that is, $W_{ii} \ne 0$, for some $i$).

Let us see which states of the pile constitute the attractor, or final class.
We have found simpler to consider the connections between two states
by means of the steepest metastable state, which is $\{22\dots 2\}$
(i.e., the one with $z(x)=2, \forall x$),
and then show which states lead to the steepest state and which ones result from it.

In fact, all states can lead to the steepest state. 
To show this, we add grains and let the sites toppling depending
on their local thresholds, but after every toppling
we assume that the maximum threshold $z_{th}(x)=2$ 
is always selected; then, we are essentially in the
same situation as in the one-dimensional BTW model:
every column in the pile grows to reach 
the steepest profile.
In this way, adding enough grains we will end in the
steepest state.
This is more easily seen applying the Abelian symmetry:
we first add grains to built the first column,
until it reaches the desired height, $h(1)=2L$,
then we add more grains and let them topple to the second column
until it reaches $h(2)=2(L-1)$, and so on.

This is enough to ensure that there is only one final class,
although the only thing we now about it is that it includes the steepest state.
If we continue with the characterization of the final class
we will simply get the attractor studied in Ref. \cite{Chua}.

On the contrary to the previous situation, not every state is reachable
from the steepest state. 
Consider first final states without zeros, i.e., $z_{f}(x)=1$ 
or $2$, only, $\forall x$.
One way to get these states is the following:
after the addition of the first grain, 
which crosses the whole pile arriving to the exit,
we fix the thresholds to the slopes of 
the desired final configuration, $z_{th}(x)=z_{f}(x)$, $\forall x$.
Then, we apply the Abelian property and 
start the toppling process of the remaining grains from the rightmost site $x=L$,
emptying out this column [from $z(L)=2$ to $z(L)=z_{f}(L)$];
after this, we take the next column to the left, $L-1$, and let 
every extra grain topple until it leaves the pile. 
Repeating the same procedure we end in the desired state
(which is therefore reachable after just one avalanche).
Basically, we are in a one-dimensional BTW-like situation again,
where the pile tends to a state $z(x)=z_{th}(x)$.

When there are zeros in the configuration,
$z(x)=0$ for some $x$, we cannot apply this trick since
thresholds are defined as larger than zero.
In fact, Chua and Christensen \cite{Chua} 
have noticed that these states do not necessarily
belong to the attractor:
they argue that zeros have to be compensated by twos 
(i.e., sites with slope $z(x)=2$);
to be precise, they show that a necessary condition to
belong to the attractor is that for each zero-slope site
in the configuration there must be at least one
two-slope site to the right, before the next zero
or before the exit.

Let us see that Chua and Christensen's condition is also sufficient
to belong to the attractor:
starting from the steepest state, the first zero appears when 
(after a number of topplings) a site with $z(x)=2$
topples (if $z_{th}(x)=1$) and the following site has $z(x+1)=1$.
Application of the rules gives then $z(x)=0$ and $z(x+1)=2$;
if $z_{th}(x+1)=2$ this site does not topple and the configuration can be metastable.
Now that a zero exists, the same process can be repeated but with the zero-slope
site 
receiving a grain;
that is, we can have $z(x-1)=2$ and $z(x)=0$,
if $z_{th}(x-1)=1$ we get $z(x-1)=0$, $z(x)=1$, and $z(x+1)=2$;
this means that the zero can move to the left, but is somehow associated to the
existence of a site with slope two, and this slope two cannot disappear 
if the zero exists. 
In order to topple, site $x+1$ would need the addition of one grain,
but grains come from the left and cannot reach $x+1$ except if the zero is removed,
i.e., any grain coming from the left would encounter the zero slope and by
the rules of the model would stick there.

In general, to get a configuration $\{ \dots 01\dots 12\dots \}$ 
from the steepest state
we go first to the configuration   $\{ \dots 11\dots 11\dots \}$ 
(the same but replacing the 0 and 2 by two 1's) 
with the thresholds equaling the slopes.
So, any new added grain will travel the whole pile up to the exit.
After this, for the site that has to have slope 2 we make its threshold 2;
an additional incoming grain will make 
the slope at this site equal to two and that of the preceding
site equal to zero; 
successive incoming grains will move the position of the zero to the desired 
site.
When there are more zero-two pairs in the final configuration
the generalization is straightforward if we start the previous procedure
from the right.

This demonstrates that the condition that each zero must have a two
to the right is a sufficient condition to belong to the attractor.
But further, the previous reasoning shows that states violating
this condition are not accessible from the steepest state,
nor from any other state which verifies the condition.
So, the condition is necessary and sufficient, and the attractor 
proposed by Chua and Christensen constitutes the only final class of
the system.

Finally, it is easy to show the existence of loops in the final class:
any state without zeros can return to the same state after the addition of one grain
if this grain travels through the whole pile and does not induce any 
other grain to topple. 
For this we need that sites with slope one have also thresholds equal to one
and sites with slope two keep their thresholds equal to two after toppling.
(The probability of this is $p^n q^{L-n}$, where $n$ is the number of sites 
with $z(x)=1$.) 
This ensures the non-periodicity of the graph and completes the demonstration
of the existence of a unique stationary distribution of state occupation.
In a case like this, the Markov chain is said to be regular.

Now that we know that there exists a stationary distribution,
how do we obtain it? 
Notice that all that we have already learn about the system has been
accomplished without explicit knowledge of the transition probabilities ${\cal W}$;
the relevant issue was if the transition was possible, $W_{ij}\ne 0$,
or not.
However, to calculate the stationary distribution and proceed further
we need the calculation of the matrix elements.

\section{Calculation of the transition probabilities}

It is possible to derive the transition probabilities
between the metastable states 
in a pile of size $L$
as a function of the transition probabilities for a size $L-1$.
Since we have fully characterized the attractor \cite{Chua},
we restrict the calculation only to these states,
for the sake of conciseness.
Note that although the number of states in the attractor 
[Eq. (\ref{Na})] becomes astronomically large for the usual 
values of $L$ in the simulations,
it constitutes a drastic reduction in front of the number
of metastable states, which is $3^L$; i.e.,
$$
{N}_A \sim 2.6^L \ll 3^L,
$$
for $L$ large.

Starting from $L=1$ there are only two possible states in the attractor,
$\{1\}$ and $\{2\}$ (state $\{0\}$
is clearly transient since the coordinate
$x=1$ corresponds already to the boundary, 
where the toppling rules for the slope are special).
We will label these two states as 1 and 2
(in this case the label is straightforward, but not for larger $L$).
Applying the rules of the model we easily get,
$$
{\cal W}^{(1)}=
\left(
\begin{array}{ll}
p & q \\
p & q \\
\end{array}
\right)
$$   
where the superindex (1) stresses that we are dealing with a system
of size $L=1$.

We can now consider $L=2$ and look at the different states there,
for instance, $\{11\}$. 
What happens when we add a grain at $x=1$?
There is a probability $p$ that the origin topples,
if not, we end in a state $\{21\}$ with a probability $q$.
If the origin topples, the grain jumps to position $x=2$ and 
there the problem reduces to an $L=1$ problem,
for which we know the transition probabilities.
In this case, we have to apply the transitions
of state 1 (which remember are defined taking into account
that one grain is added to this state at its leftmost position, $x=2$ now);
as these transition probabilities are
$W^{(1)}_{11}$ and $W^{(1)}_{12}$,
we have for the state $\{11\}$ the probabilities:
\begin{equation}
W^{(2)}_{ \{11\} \{z(1) z(2)\} } =
\left\{
\begin{array}{ll}
q &\mbox{ to go to state } \{z(1) z(2)\} = \{21\},\\
p W^{(1)}_{11}=p^2 &\mbox{ to return to } \{11\},\\
p W^{(1)}_{12}=pq  &\mbox{ to go to } \{02\},\\
0 &\mbox{ for any other final state.}\\
\end{array}
\right.
\end{equation}

This simple example shows how to reduce the problem from $L$ to $L-1$.
In general, we will refer to the $L$ system as the pile, or just the system,
and the $L-1$ pile, defined by $x=2,3 \dots L$, will be the subsystem or subpile.
In the same way we will talk about states of the system and about substates
when referring to the subsystem.

Although the previous case illustrates the basic idea, 
there appears an extra complication 
if the height at the origin, $h(1)$, is larger, 
which is that the origin can topple more than once 
if the avalanche in the $L-1$ subpile is big
enough to leave the origin with a too large slope. 
In this case one can apply the Abelian property: let us topple first
the subpile (just using the transition probabilities 
${\cal W}^{(L-1)}$ that we know)
and then, at the end, let the origin topple.
Of course, this gives rise to an iterative process, 
where the procedure has to be
applied as many times as the origin topples, which is $h_I(1)-h_F(1)+1$,
$I$ and $F$ referring to the initial and final states.
The sequence is: first the origin topples, then, the $L-1$ subpile
to reach equilibrium, then the origin again (if needed), then the subpile,
and so on.

Applying the previous argument to a general system of size $L$ 
one can find the rules for the transition probabilities.
It is convenient to define a variable $Q$ as 
$$
Q \equiv h(1)-L=\sum_{x=1}^L z(x)-L;
$$
we have $Q=0,1 \dots L$ in the attractor \cite{Chua}.
The equations for the elements of ${\cal W}^{(L)}$ will
depend on $\Delta Q=Q_I-Q_F$, that is, the difference of $Q$ between the initial
state and the final state (defined here in the opposite way as usual);
this is so because the number of topplings at $x=1$ is $\Delta Q +1$.
The rules are given below and refer to the transition probabilities 
between an initial state $I$
with a value of $z (1)=z_I$ and $L-1$ subpile state $i$
and a final state $F$ with $z (1)=z_F$ and substate $f$,
that is, a transition $(z_I,i) \rightarrow (z_F,f)$.
Note how an $L-$state is completely characterized
by the value of $z (1)$ and the state (substate) of the $L-1$ subsystem.
As with every toppling of the origin $Q$ decreases in one unit,
we will use the following relation to calculate $z(1)$,
$$
z(1)=Q-Q'+1,
$$
where $Q'$ refers to the $Q$-value
of the $L-1$ subsystem.
In general, $Q'_s$ will denote the $Q$-value of substate $s$ 
in the $L-1$ subsystem.
$T(z)$ will give the probability
that a site topples for a given value of $z$ 
[$0$, $p$, $1$ for $z\le 1$, $2$, $\ge 3$ 
respectively, see Eq. (\ref{alternative})].
Therefore, the argument of $T$ is the value of $z(1)$ calculated 
after a number of topplings.
With all these definitions the rules turn out to be,
\begin{equation}
W^{(L)}_{(z_I,i) (z_F,f)}= \left\{
\begin{array}{ll}
= 0  & \mbox{ if } \Delta Q \le -2,
\\
= [1-T(z_I+1)] \delta_{i f} 
& \mbox{ if } \Delta Q=-1, 
\\
=T(z_I+1) W^{(L-1)}_{i f } [1-T(Q_I-Q'_f+1)] 
& \mbox{ if } \Delta Q=0,
\\
=T(z_I+1) \sum_j W^{(L-1)}_{i j } T(Q_I-Q'_j+1) W^{(L-1)}_{j f}
[1-T(Q_I-Q'_f)]  
&\mbox{ if } \Delta Q=1,
\\
=T(z_I+1) \sum_j W^{(L-1)}_{i j } T(Q_I-Q'_j+1) 
\\
\sum_k W^{(L-1)}_{j k}T(Q_I-Q'_k)
              W^{(L-1)}_{k f}[1-T(Q_I-Q'_f-1)] 
& \mbox{ if } \Delta Q=2.
\\
\end{array}\\
\right.
\end{equation}
The case $\Delta Q \le -2$ is impossible since we would need 
adding more than one grain at $x=1$ (and that some of them would not topple) 
to reach the corresponding value of $Q$.
For $\Delta Q=-1 $ the only possibility is that the origin does
not topple, then $z_I$ increases in one unit and has to be stable,
and the substate does not change.
For the rest of cases we can write
\begin{equation}
\begin{array}{ll}
W^{(L)}_{(z_I,i) (z_F,f)}
=T(z_I+1) \sum_j W^{(L-1)}_{i j } T(Q_I-Q'_j+1) 
\sum_k W^{(L-1)}_{j k}T(Q_I-Q'_k) \\
       \sum_l W^{(L-1)}_{k l} T(Q_I-Q'_l-1)
       \cdots 
       \sum_v W^{(L-1)}_{u v} T(Q_I-Q'_v-\Delta Q+2) \\
              W^{(L-1)}_{v f} 
        [1-T(Q_I-Q'_f-(\Delta Q-1))]  
       & \mbox{ if } \Delta Q \ge 0. 
\end{array}
\end{equation}
\noindent
The general idea is that there is a probability $T(z_I+1)$ that
the origin topples after the addition of one grain, 
then, there is a probability $W^{(L-1)}_{i j}$ to go to a substate $j$;
with this substate there is a probability 
that the origin topples given by $T$ again,
and then, one goes from $j$ to $k$, from here to $l$, etc.,
following all the possible paths that end in $f$,
whose origin has a probability $1-T(z_F)$ to be stable,
with $z_F=Q_F-Q'_f+1=Q_I-Q'_f-\Delta Q+1$.

The previous equations for the elements of ${\cal W}^{(L)}$ 
look like a matrix product but with different matrices 
and a different number of factors for different components.
Nevertheless in matrix form they can be written as
%
$$
W^{(L)}_{(z_I,i) (z_F,f)}=
T(z_I+1)[1-T(z_F)] 
\left[ \left( \prod_{Q=Q_I}^{Q_F+1} 
{{\cal W}^{(L-1)}} \cdot {\cal T} _{Q} \right)
\cdot {{\cal W}^{(L-1)}} \right]_{i f}
\mbox { if } \Delta Q \ge 0,
$$
where the index $Q$ is assumed to decrease in the product,
$[\ ]_{i f}$ denotes that we take the element 
of the matrix in the $i$ file and $f$ row and
${\cal T}$ is a diagonal matrix whose elements 
can only be $1,p$, or $0$,
more precisely,
$$
\left[ {\cal T}_{Q }\right]_{j k}=T(Q-Q'_j+1) \delta_{jk}.
$$
We stress that the rules are valid for any pair of metastable states,
although we will concentrate on states in the attractor.

These equations for $L=2$ yield
\begin{equation}
{\cal W}^{(2)}=
\left(
\begin{array}{lllll}
0 & 0 & 0 & 1 & 0 \\
pq & p^2 & q & 0 & 0  \\
p^2q & p^3 & pq & q & 0 \\
p^3q & p^4 & p^2q & pq & q \\
p^3q(1+q) & p^4(1+q) & p^2q(1+q) & pq(1+q) & q^2 \\
\end{array}
\right)
\nonumber
\end{equation}
where states are ordered as $\{02\}, \{11\}, \{21\}, \{12\}, \{22\}$.
Iterating the rules it is possible, although laborious, 
to generate the matrices ${\cal W}^{(L)}$
for successive $L$. 
Although the matrix for $L=2$ looks simple, 
the corresponding matrices as $L$ increases
are getting more and more complicated.


Figure 1 shows, for $L=7$ and 8, and $p=1/2$, 
the probability density $H(K)$ 
of the number of nonzero elements for each file of $\cal W$,
that is,
the distribution of the number of states
directly accessible for a state in the attractor,
or, in the language of networks, the out-degree distribution
of the phase space.
There seems to be two kinds of states, one group has few connections
and the other one a large number of them;
nevertheless the system size is too small to be conclusive.
In contrast, the in-degree distribution
(not shown in the plot) looks rather uniform.
(We will always use the letter $H$ to denote probability densities,
although it will correspond to different functions depending
on the argument.)

In Fig. 2 we illustrate the enormous variation in the values of 
the transition probabilities:
the probability density $H(W)$ that $W_{ij}$ takes a certain value $W$ 
is shown for $L=7$
and $p=1/2$, spanning about 16 orders of magnitude.
A power law with exponent minus one approximates well this behavior.
Curiously, a similar result 
has been found in networks describing 
correlations between earthquakes \cite{Paczuski2003b}.


\section{Obtaining of the stationary occupation distribution}


To get the evolution of the system one has to start with a 
distribution of initial states ${ P}_0$,
which remember has to be an ${N}_A-$dimensional vector 
[see Eq. (\ref{Na})],
giving the occupation probability of any of the ${N}_A$
states in the attractor.
This initial distribution can be a delta 
[for instance, for $L=2$, starting always  
with $\{11\}$, i.e., ${ P}_0=(0, 1, 0, 0, 0)$] or not.
The distribution of states after the first (slow) time step,
i.e., after the addition of just one grain,
is obtained as $P_1 = { P}_0 \cdot {\cal W}$.
To obtain the distribution of states 
in the next time step we have to multiply again by ${\cal W}$ and so on;
the powers of ${\cal W}$ give therefore the evolution
of the system.

Let us call the vector representing the stationary distribution ${D}$, 
which of course is also a vector on ${N}_A$ dimensions,
with each component $D(s)$ giving the probability that after a long enough time
the system is in state $s$.
The evolution of the occupation probabilities in the attractor
is also obtained multiplying the row vector ${D}$
by the matrix ${\cal W}$, but as ${D}$ is the stationary distribution, it must be
invariant under such operation, i.e., ${D}={D} \cdot {\cal W}$,
which simply means that ${D}$ is a left eigenvector of ${\cal W}$ with
eigenvalue equal to one. 
Regularity (which was demonstrated in Sec. III)
ensures that this eigenvector is unique.

Moreover, a direct consequence of regularity,
which is also a sufficient condition for it, is that
\begin{equation}
{\cal M} \equiv 
\lim_{n \rightarrow \infty} {\cal W}^n =
\left(
      \begin{array}{c}
       D\\
       D\\
       . \\
       . \\
       . \\
       D
      \end{array}\\
\right)
\end{equation}
which means that the transition matrix corresponding to $n$ steps
has all its files equal to the stationary distribution $D$,
asymptotically.
Indeed, it is trivial to show that for any distribution $P$,
with $\sum_i P(i)=1$, we have $P \cdot {\cal M}=D$.
On the opposite side, if any $P$ tends to $D$ we can take
$P=(1,0 \dots 0)$ to get the first file of matrix $\cal M$
(by multiplication), which must be equal to $D$,
and the same can be done for any other file of $\cal M$.

If we consider the case $L=2$ we realize that
${\cal M}={\cal W}^3$, that is,
we obtain a matrix whose files are all the same;
this implies that 
the asymptotic result is reached in just three time steps,
since ${\cal M} ={\cal W} \cdot {\cal M} ={\cal W}^2 \cdot {\cal M}$, etc.
But further, it turns out that the stationary distribution $D$,
given by the files of $\cal M$,
is the last file of $\cal W$, if the states are ordered by
increasing $Q$, so,
$$
D=
\left(
\begin{array}{lllll}
p^3q(1+q), & p^4(1+q), & p^2q(1+q), & pq(1+q), & q^2 
\end{array}
\right);
$$
that is, the transition probabilities of the steepest state $\{22\}$
give the occupation probabilities in the stationary regime.
In other words, the unique eigenvector of $\cal W$ with eigenvalue equal to one
is just its last file.

We have verified that this result is general for larger $L$,
although the necessary number of powers increases with $L$
(note that for $L=1$ this is already accomplished at the first time step).
Dhar \cite{Dhar2003} has beautifully demonstrated this result 
using the properties of an operator algebra. 
The idea behind this is simple:
we can realize that it is equivalent to add $L(L+1)$ grains
to the flattest state $\{00\dots 0\}$ (which is not in the
attractor) than to add just one grain to the steepest state
$\{22\dots 2\}$. Let us see why.
The number of grains which separate both profiles
is precisely $L(L+1)$, so, by application of the Abelian 
property, we add this number of grains to $\{00\dots 0\}$
and let them topple to reach the profile corresponding to
the steepest state (the probability of this toppling process
is exactly one); after reaching this state we can continue the toppling
process, but we are already in the same situation that results
from adding just one grain to the steepest state.
So we get the same configurations with the same probabilities in both
cases.
In fact this result is not only true for the flattest state, but
for any other state, the only difference is that we will have some
extra grains: no problem, they topple until they leave the pile.
Therefore we can write
$$
(0\dots 0,1) \cdot {\cal W} = P_0 \cdot {\cal W}^{L(L+1)},
$$
$\forall P_0=(0\dots 0,1,0\dots 0)$.
(This expresses something that we already proved in Sec. III,
which is that the steepest state is reachable from any configuration,
just adding enough grains.)
Note now that the addition of one extra grain changes nothing in each case,
this grain will topple until the exit, so
$$
P_0 \cdot {\cal W}^{L(L+1)}=P_0 \cdot {\cal W}^{L(L+1)+1}.
$$
As this equality holds for any vector of the basis,
we can write
$$
{\cal W}^{L(L+1)}={\cal W}^{L(L+1)+1}={\cal M},
$$
and so, from the first equation we get
$$
(0\dots 0,1) \cdot {\cal W} = P_0 \cdot {\cal M}= D, \ \forall P_0,
$$
which means that, indeed, the transitions from the steepest state
coincide with the stationary occupation distribution.
Dhar has also noted that a more restrictive condition holds 
for the states in the attractor. 
The state there with less grains is $\{11\dots 1\}$;
the difference in number of grains between the steepest state
and this one is $L(L+1)/2$, which can replace the previous
value $L(L+1)$, for any state in the attractor.

In Fig. 2 we also include the probability density $H(D)$ that $D(s)$
takes a given value for $L=7$ and $p=1/2$, 
showing a behavior very similar to the density
of transition probabilities;
this is a broad distribution across 13 orders of magnitude,
close to a power law with exponent minus one.
This means that the occupation of the phase space
(i.e., the space of all possible configurations)
is enormously heterogeneous,
at variance with the BTW model \cite{Dhar}.

Figure 3 shows $D(s)$ as a function of $s$, where the states $s$
are ordered in terms of decreasing $D(s)$.
In fact, the form of $D(s)$ in this plot is related to $H(D)$,
just by identifying $s/N_A$ as the probability that the occupation
probability is larger than (or equal to) a certain value $D(s)$, that is, 
as the survivor function of the random variable $D(s)$.
Therefore, the density $H(D)$ will be (as usual) the derivative
of this survivor function, multiplied by  $-1$, or
$$
H(D)=-\frac 1 N_A \left( \frac {d D(s)}{ds}\right)^{-1}.
$$
A power law with minus one exponent for $H(D)$ yields an
exponentially decreasing $D(s)$, in agreement with the plot.

\section{Calculation of the distributions of mean slopes and
avalanche sizes}

From the values of the stationary distribution of the occupation of the states 
and their transition probabilities,
$D(s)$ and ${\cal W}$,
it is possible to calculate many things in the stationary state,
(i.e., in the attractor).
The first one is the (stationary) distribution of $Q$, $f(Q)$, 
$$
    f(Q)
    =\sum_{\forall s \ s.t. \  Q_s=Q} D(s) 
    =\sum_{\forall s }  D(s) \delta_{Q_s Q}. 
$$
For example, for $L=2$ we get 
$$
f(Q)=
\left(
\begin{array}{lll}
p^3(1+q), &  pq(1+p)(1+q), & q^2 
\end{array}
\right),
$$
for $Q=0, 1, 2$.

The distribution $f(Q)$ is in fact the distribution 
of heights at the origin, $f(h(1))$, 
and it is also directly related to the distribution
of mean slopes, $f(\bar z)$, 
with $\bar z \equiv \sum_{x=1}^L z(x)/L=h(1)/L$, since
$Q=h(1)-L=L(\bar z-1)$.
Consequently, $f(Q)$ defines the active zone width,
which can be obtained as the standard deviation of this distribution.
(For simplicity, we have used the same symbol $f$ for all the distributions,
although obviously they are not the same function.)

The distribution for the value of 
$(Q-\langle Q \rangle)/L^{\chi}=
(h(1)-\langle h(1) \rangle)/L^{\chi}=
(\bar z-\langle \bar z \rangle)L^{1-\chi}$
is displayed in Fig. 4 for several values of $L$.
For $\chi$, we take the value proposed in Ref. \cite{Corral99}, 
$\chi \simeq  0.24$.
Note how all the discrete distributions collapse onto a single continuous
curve under rescaling, which is close to Gaussian,
though slightly skewed.
These exact results for small $L$ are in total agreement
with the findings of computer simulations.

The avalanche size distribution $f(S)$ (in the attractor) 
is not difficult to calculate 
knowing $D(s)$ and $\cal W$.
We can write
$$
f(S) = \sum_{\forall  s} p(S / s) D(s),
$$
where $S$ is the avalanche size, $s$ the state of the pile,
and $p(S /s) $ is the conditional probability of having an avalanche of
size $S$ starting from a state $s$.
For this term we have,
$$
p(S/s) 
       = \sum_{\forall j \ s.t. \  S_{sj}=S} W_{s j}
       = \sum_{\forall j} W_{s j} \delta_{S_{sj} S},
$$
where $S_{sj}$ is the size of the avalanche triggered in the 
transition from $s$ to $j$ and can be calculated as
$$
S_{sj}=\sum_{x=1}^{L}(h_s(x)-h_j(x))(L-x+1)+L,
$$
which is essentially the profile difference times the distance to the exit,
plus the contribution of the added grain.
Therefore we have,
$$
f(S) = \sum_{\forall s} D(s) 
\sum_{\forall j} W_{s j} \delta_{S_{sj} S}.
$$

The corresponding distribution calculated in this manner 
for $L=8$ and $p=1/2$ appears in Fig. 5,
where it is compared with the result obtained from computer simulations.
Note how even for such a small system there are avalanches with probability
smaller than $10^{-22}$.

%

\section{Discussion}

Just to summarize, we have shown that the conditions about the states
put forward in Ref. \cite{Chua} are necessary and sufficient to
define the attractor, which moreover is unique.
In addition, its nonperiodical character allows the existence of a single
occupation distribution in the stationary limit.
The Abelian property enables the calculation of the transition
probabilities between the different states, 
just from the decomposition of a system of size $L$ into its leftmost column
and an $L-1$ subsystem and by using an iterative toppling procedure;
this allows to explore the network of connections in phase space.
Unexpectedly, the stationary occupation distribution turns out to be the last file
of the transition matrix, i.e., that corresponding to the transitions
of the steepest state.
Both the transitions between states and their occupations
can take values in an extremely large range of probabilities,
setting a clear difference with other SOC systems.
These calculations are exact for the system sizes involved,
and could be performed for any $L$, in principle.
In practice, we have strong limitations, as it is explained below.
Finally, with these quantities we can also derive the form of the
fluctuations of the profile and the avalanche size distribution,
for the corresponding value of $L$.

In fact, the knowledge of the transition probability matrix and the stationary
distribution allows the calculation of any property related
to the profile of the pile, as the
ones we have just mentioned
or the dissipated-energy distribution.
But there are other properties that do not only depend on the 
profile but also on the dynamics which leads from one profile to another,
for instance, the avalanche duration: knowing the profiles is not enough
to calculate this quantity; in fact, with the same initial and final
states the duration is not uniquely defined, i.e., there are many ways
to end in the same state with a different number of avalanche time steps,
depending on the actual sequence of topplings. 
So, despite the usefulness of the Abelian property, it does not allow 
to calculate dynamical magnitudes. 
This could be solved in principle with a parallel calculation
stating the probability that a given transition would involve a given
time; nevertheless, it would be complicated.

Also, the profiles, that is, the configurations defined in terms
of slopes (or heights), do not retain any information regarding
individual grains.
Grains are essentially treated as undistinguishable
whereas for the calculation of transit times
or flight lengths one needs distinguishable particles.
This is another limitation of the current method
difficult to overcome.

On the other hand,
the equations obtained for the transition probabilities have the advantage
that one does not have to simulate the system, and therefore one obtains
exact probability distributions.
However, the enormous number of states in the attractor
(which increases as $2.6^L$) makes impossible any calculation by hand beyond
the smallest values of $L$.
Symbolic computer calculations, performed by MAPLE or similar programs,
have also severe limitations of size,
and even with numeric computations the system size is limited to about
$L \le 10$ (for $L=10$ there are more than $10^4$ recurrent states, 
which would require $10^8$ matrix elements).
Also, the number of different matrix products is large.
Although the results we obtain for small systems
are interesting enough and representative of the complexity
of the system, it would be nice to have access to supercomputers
to increase the capabilities to manage larger system sizes.

To conclude,
it is interesting to point out that having exact analytical results
for some problem is not a synonymous of understanding;
in this case we can obtain exact expressions for the transition probabilities,
the stationary distribution, or the avalanche size distribution,
but from these formulas still it is not clear which are the properties of the
system for large $L$. 
For instance, for the avalanche size distribution we can get very complicated
exact equations, but we cannot show that they tend to a power law in the
asymptotic case.
Nevertheless, in our opinion, the results in this paper imply a
remarkable progress in our picture of complex systems.

\section{Acknowledgements}

I am grateful to M. Bogu\~n\'a for discussions; 
he also was, as far as I know,
the first one to study the transition probabilities between
the different states of the Oslo model, long time ago.
I am also indebted to K. Christensen and D. Dhar for the
information about their work previous to publication
and to the latter for his comment on this manuscript.
These discussions and many others were possible thanks to the
organizers of the Symposium 
{ \it Complexity and Criticality},
held in Copenhagen in memoriam of Per Bak.
Finally, I thank the Spanish MCyT the creation of the Ram\'on y Cajal 
program.

%
\begin{figure}
\epsfxsize=6truein 
\hskip 0.15truein\epsffile{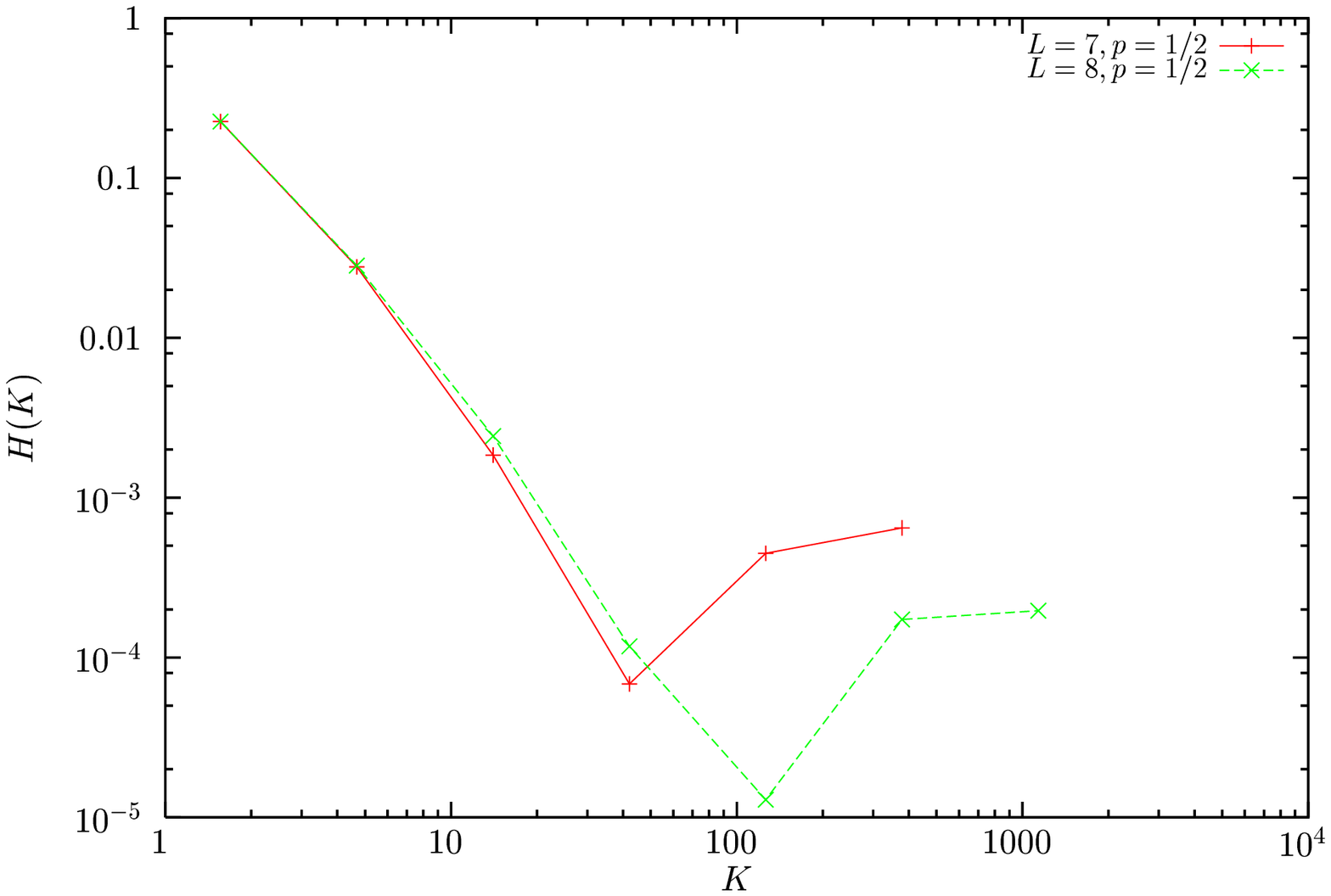} 
\caption{ 
Probability density that the 
number of states directly accessible in phase space
(or out-degree distribution)
for a state in the attractor takes a value equal to $K$,
for $L=7$ and 8 and $p=1/2$.
For $K$ not very large, $H(K)$ could be a power law,
but larger values of $L$ are needed to be more sure.
The histogram is calculated with exponentially increasing bins.
}
\end{figure}
%

%
\begin{figure}
\epsfxsize=6truein 
\hskip 0.15truein\epsffile{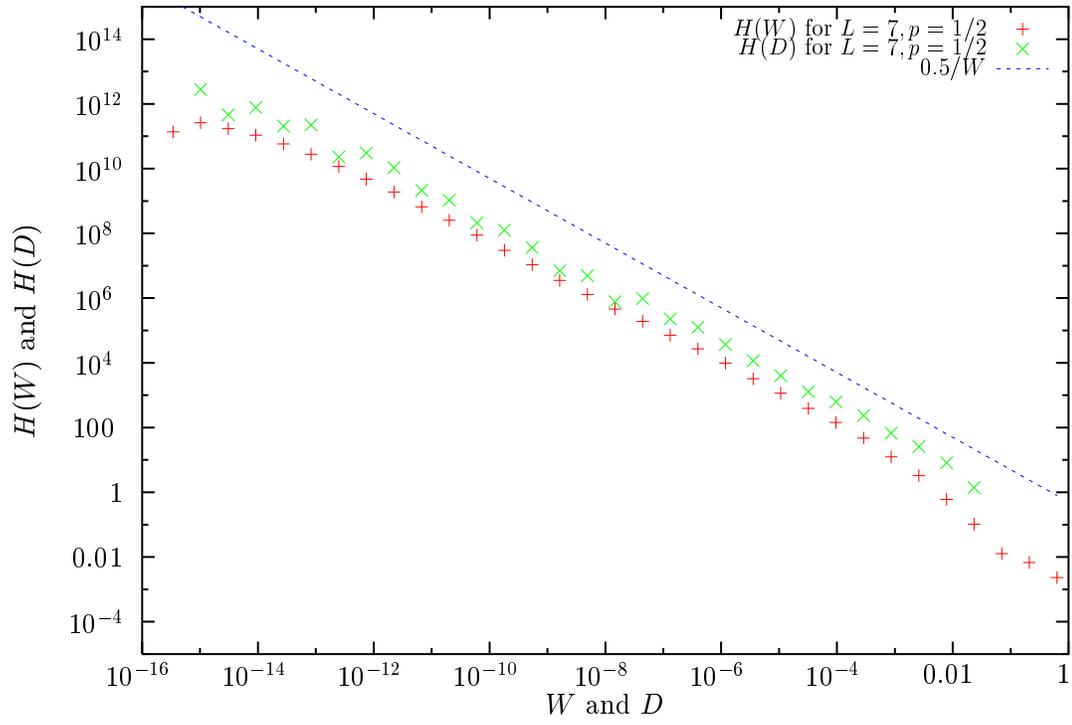} 
\caption{ 
Probability densities $H(W)$ and $H(D)$
that some transition probability takes a value $W$, and
that some state $s$ has stationary occupation probability $D(s)=D$, 
for $L=7$ and $p=1/2$.
\label{DW}
}
\end{figure}
%

%
\begin{figure}
\epsfxsize=6truein 
\hskip 0.15truein\epsffile{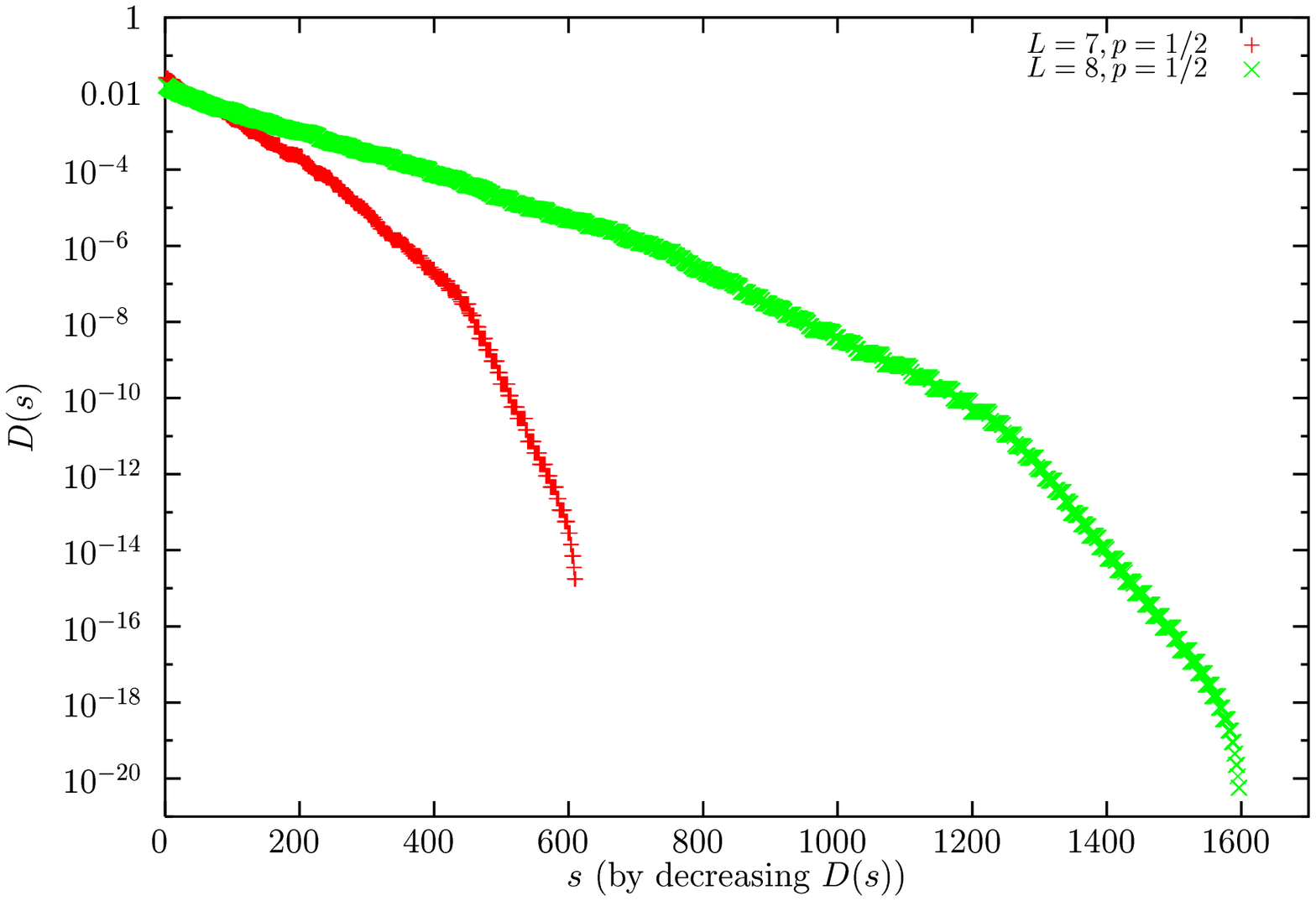} 
\caption{ 
Stationary probability of occupation $D(s)$ for each state for $L=7$
and 8,
and $p=1/2$.
The states are ordered in decreasing probability.
Straight lines would indicate an exponential decay with $s$.
\label{Poc}
}
\end{figure}
%

%
\begin{figure}
\epsfxsize=6truein 
\hskip 0.15truein\epsffile{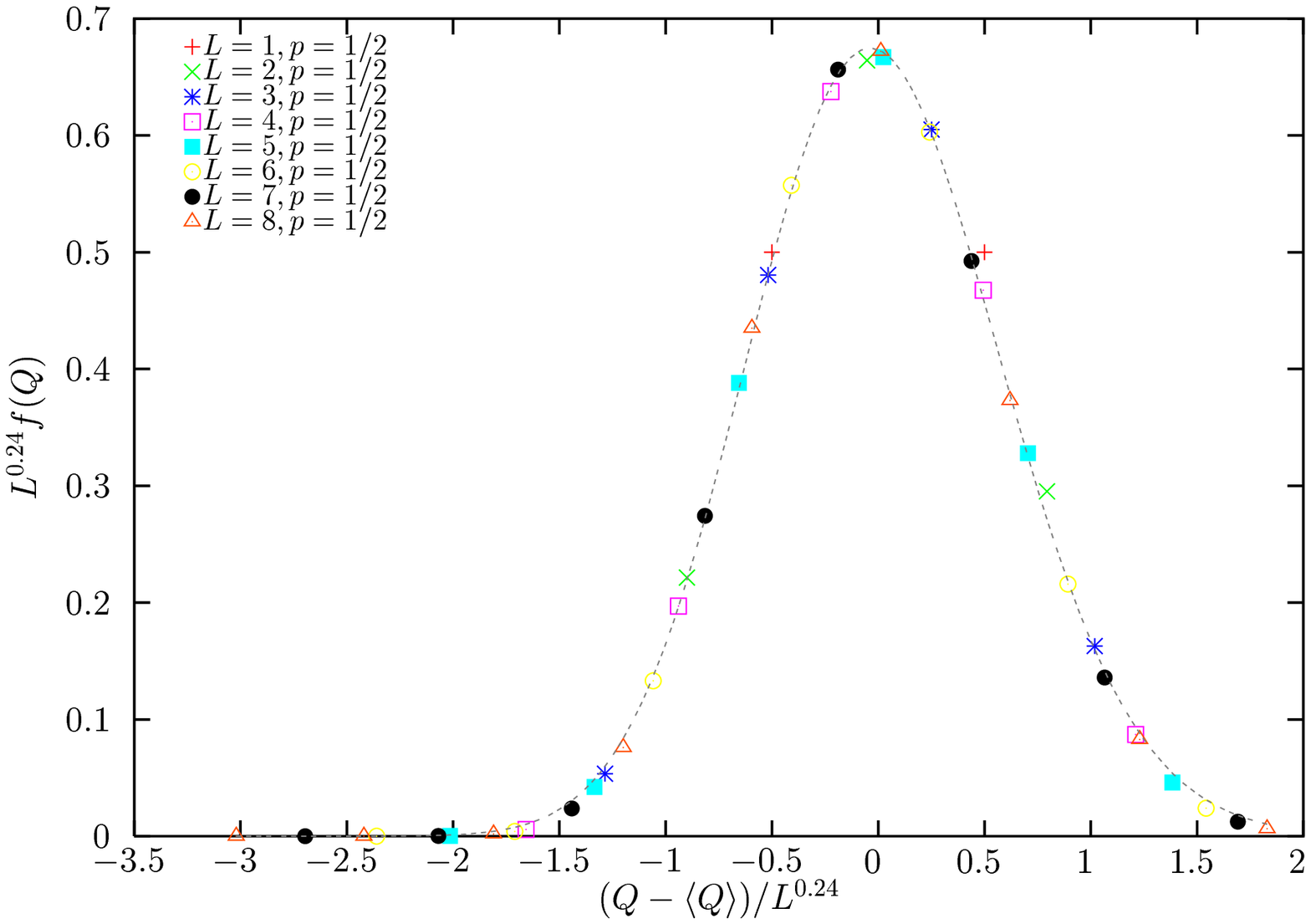} 
\caption{
Stationary distribution of $Q$ (or of heights at the top, $h(1)$), 
centered by the mean and scaled by $L^{0.24}$.
System size ranges from $L=1$ to $L=8$, and $p=1/2$. 
Notice how all the discrete distributions for each system size conspire
to give a smooth curve,
which is close to Gaussian, although slightly skewed.
\label{Dqsca}
}
\end{figure}
%

\begin{figure}
\epsfxsize=6truein 
\hskip 0.15truein\epsffile{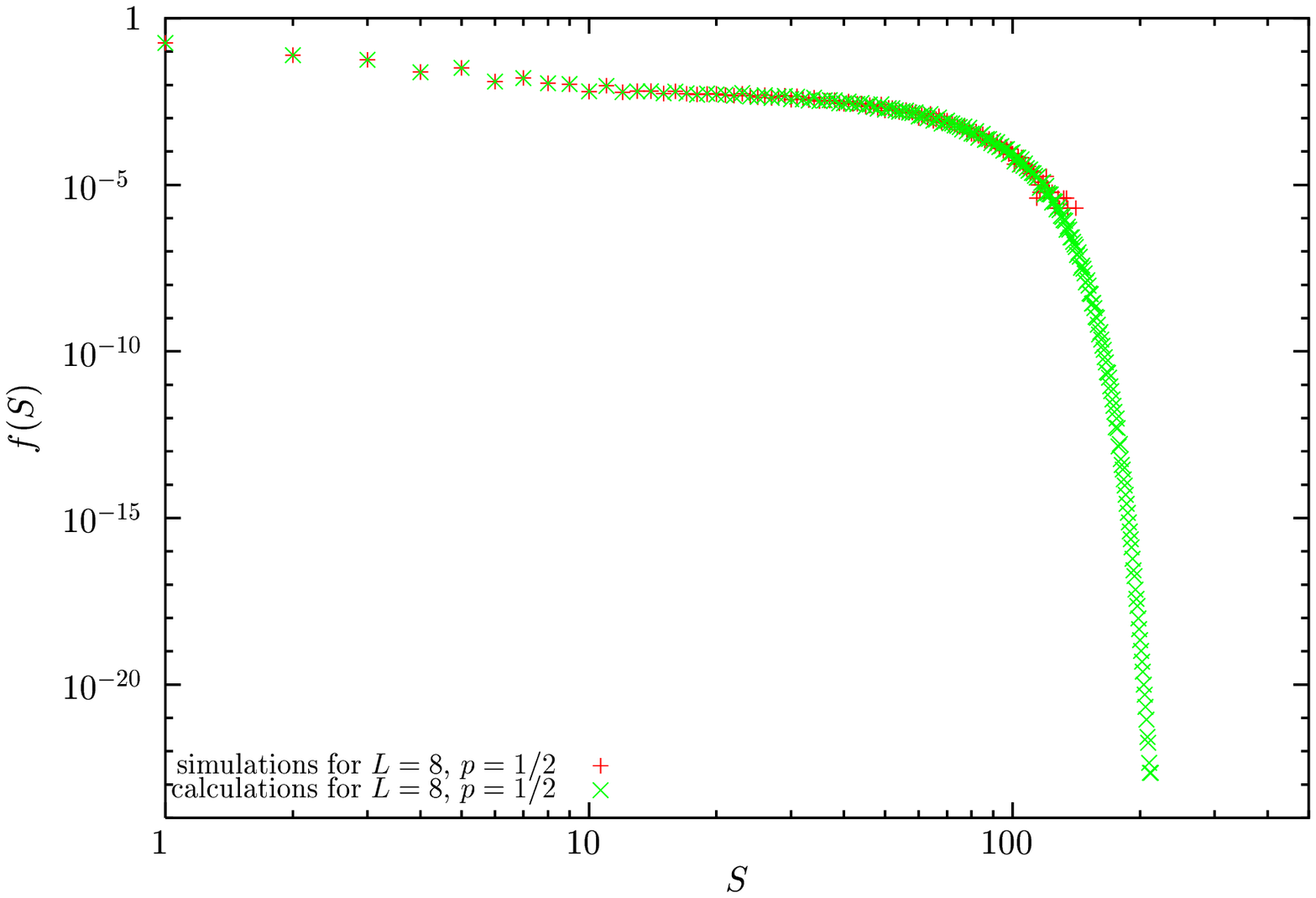} 
\caption{ 
Stationary distribution of avalanche sizes for $L=8$ and $p=1/2$
from our exact procedure and from
simulations of the Oslo model.
\label{Dsize}
}
\end{figure}
%

%
%

%
%


\begin{references}
\bibitem{negative}
If we took as an initial condition a configuration
with negative slopes, the rules would lead the system,
after some time, to only positive slopes.
\end{references}
\end{document}